\documentclass[10pt,conference]{IEEEtran}
\IEEEoverridecommandlockouts
\usepackage{listings}
\usepackage{xcolor}
\definecolor{lightgray}{gray}{0.75}

\usepackage{booktabs}
\usepackage{multirow}
\usepackage{colortbl}
\usepackage[table]{xcolor}
\usepackage{longtable}
\usepackage{url}

\usepackage{hyperref}
\usepackage{subcaption}

\usepackage{authblk}

\usepackage{cite}
\usepackage{amsmath,amssymb,amsfonts}
\usepackage{algorithmic}
\usepackage{graphicx}
\usepackage{textcomp}
\usepackage{xcolor}
\def\BibTeX{{\rm B\kern-.05em{\sc i\kern-.025em b}\kern-.08em
    T\kern-.1667em\lower.7ex\hbox{E}\kern-.125emX}}

\begin{document}

\title{Towards Context-aware Mobile Privacy Notice: Implementation of A Deployable Contextual Privacy Policies Generator}


\author[1,*]{Haochen Gong}
\author[1,2*]{Zhen Tao}
\author[1,2,$^\dagger$]{Shidong Pan}
\author[1,2]{Zhenchang Xing}
\author[1]{Xiaoyu Sun}
\affil[1]{\textit{School of Computing, Australian National University}}
\affil[2]{\textit{CSIRO's Data61}}
\affil[*]{\small Equal contribution  $^\dagger$Project Lead (shidong.pan@data61.csiro.au)}

\maketitle

\begin{abstract}
Lengthy and legally phrased privacy policies impede users' understanding of how mobile applications collect and process personal data. Prior work proposed Contextual Privacy Policies (CPPs) for mobile apps to display shorter policy snippets only in the corresponding user interface contexts, but the pipeline could not be deployable in real-world mobile environments. In this paper, we present \textbf{\textsc{PrivScan}}, the first deployable CPP Software Development Kit (SDK) for Android. It captures live app screenshots to identify GUI elements associated with types of personal data and displays CPPs in a concise, user-facing format. We provide a lightweight floating button that offers low-friction, on-demand control. The architecture leverages remote deployment to decouple the multimodal backend pipeline from a mobile client comprising five modular components, thereby reducing on-device resource demands and easing cross-platform portability. A feasibility-oriented evaluation shows an average execution time of 9.15\,s, demonstrating the practicality of our approach. The source code of \textbf{\textsc{PrivScan}} is available at~\url{https://github.com/buyanghc/PrivScan} and the demo video can be found at~\url{https://www.youtube.com/watch?v=ck-25otfyHc}.
\end{abstract}


\section{Introduction}
\label{sec_intro}

Mobile applications are an integral part of people’s digital lives and collect a significant amount of personal data from users. In response to growing privacy concerns regarding data collection practices, mobile application companies are required to provide users with privacy notices, e.g., privacy policies, that describe their privacy practices. Regulations, such as the European Union's General Data Protection Regulation (GDPR)~\cite{GDPR} and California’s Consumer Privacy Act (CCPA)~\cite{ccpa2018}, require developers to provide understandable privacy notices to inform users about how their data is collected and used. Privacy policies have become the most common means of data practices disclosure. Mobile app platforms, such as Google Play and Apple App Store, also require applications on the platform to provide privacy policy links~\cite{DeveloperPolicy, AppGuidelines}.

Although mobile applications require users to accept their privacy policies, these policies are often so lengthy that most users bypass them to gain immediate access to the app’s functionality~\cite{wagner2022privacy}. 
When users attempt to read the privacy policies, they often encounter complex legal language that is difficult for non-experts to understand~\cite{mcdonald2009comparative}.
Similarly, developers also face difficulties in adequately interpreting and understanding privacy policies due to limited legal knowledge, ultimately affecting their ability to implement compliant software behaviors~\cite{tao2025privacy}.
Moreover, privacy policies are isolated from actual usage contexts, making it difficult for users to relate the abstract descriptions to their real-time interactions \cite{shvartzshnaider2019going}, thereby failing to adequately engage user awareness.

In response to these challenges, efforts have been made to design more usable privacy notices tied to their immediate context. Both iOS and Android introduce intrinsic permission popup mechanisms~\cite{RequestAuthorizationApple, PermissionsAndroid}, which notify users the first time the app attempts to obtain sensitive device permissions. Code-based IDE plugins, such as Matcha~\cite{li2024matcha}, have also been proposed to guide developers in creating accurate privacy nutrition labels for their apps.
However, these approaches often fall short in efficacy, as they do not present contextually relevant statements in the privacy policy to users.

The concept of Contextual Privacy Policies (CPPs) has been proposed as a user-centered alternative~\cite{ windl2022automating, pan2024new} that aims to extract shorter and more digestible segments from the traditional privacy policy, and dynamically display them in relevant contexts on the user interface (UI). CPPs help users become aware of potential data collection through UI elements and understand the relevant privacy disclosures in privacy policies. Such situated awareness bridges the gap between intention and action, facilitating informed consent in context.


In this paper, we present \textbf{\textsc{PrivScan}}, the first CPP Software Development Kit (SDK) for mobile applications. The underlying approach of this tool was proposed in our previous work~\cite{pan2024new}, known as the \textbf{\textsc{SEEPRIVACY}} framework. It leverages computer vision and natural language processing techniques to analyze user interface screenshots alongside corresponding privacy policy texts. It maps UI elements to specific personal data type and displays relevant policy policy snippets in context to help users understand the potential data practices associated with their interactions. While this framework marks a significant step toward contextualizing privacy, it remains at an experimental stage, lacking the capability for deployment in real-world mobile application scenarios, which significantly limits its practical applicability. 

\begin{figure} [t!]
\begin{subfigure}{.32\linewidth}
  \centering
  \includegraphics[width=.99\linewidth]{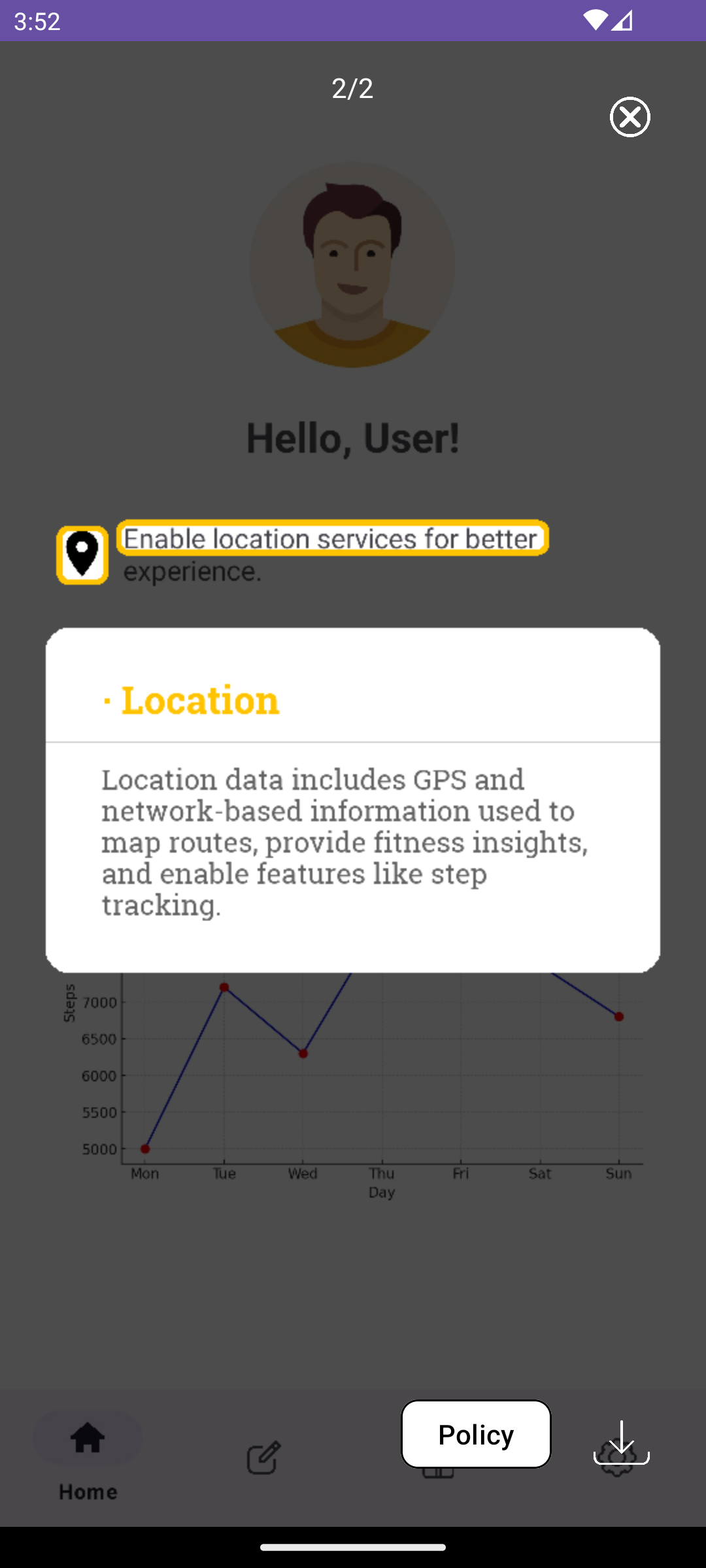}
  \caption[]{Location}
  \label{fig_1}
\end{subfigure}%
\hspace{0.7pt}
\hfill
\begin{subfigure}{.32\linewidth}
  \centering
  \includegraphics[width=.99\linewidth]{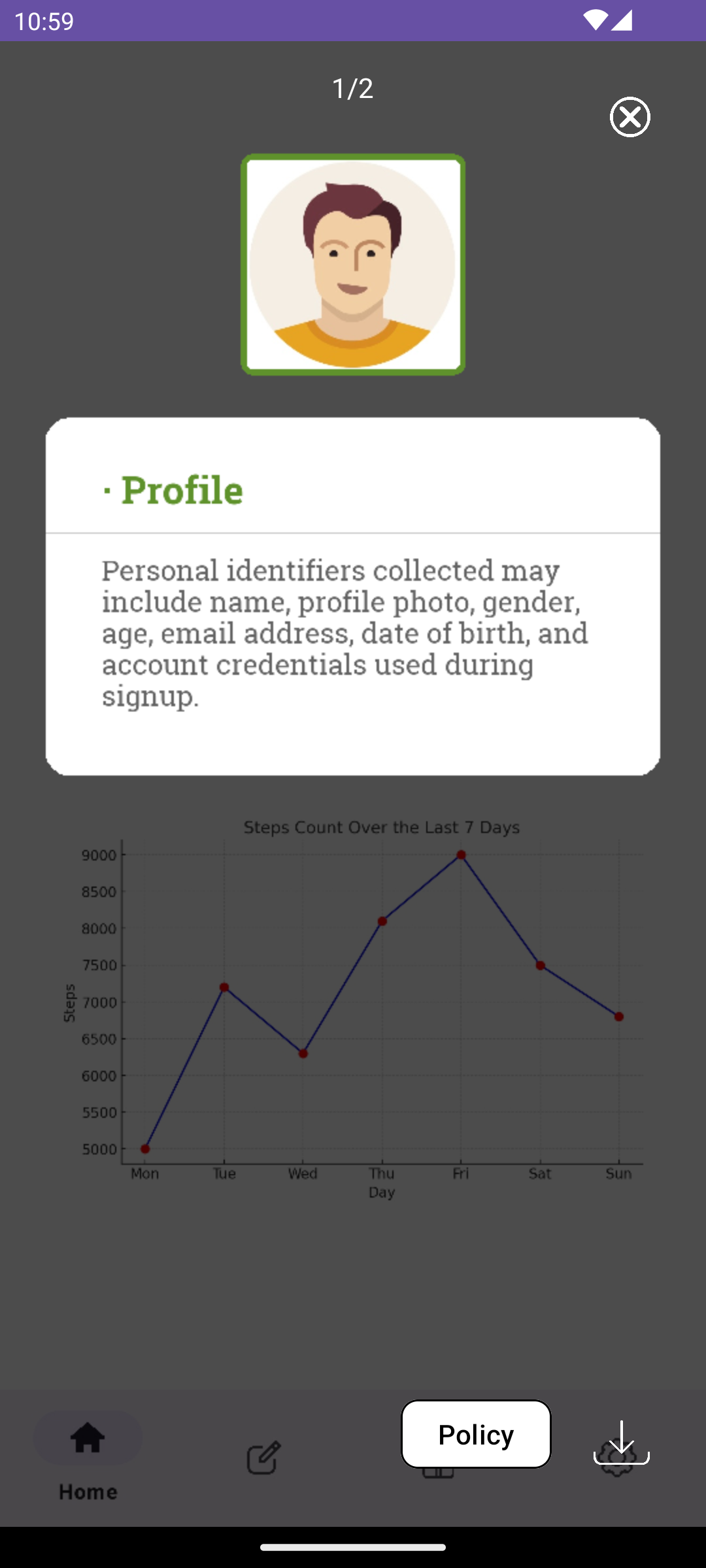}
  \caption{Profile}
  \label{fig_2}
\end{subfigure}
\hfill
\begin{subfigure}{.32\linewidth}
  \centering
  \includegraphics[width=.99\linewidth]{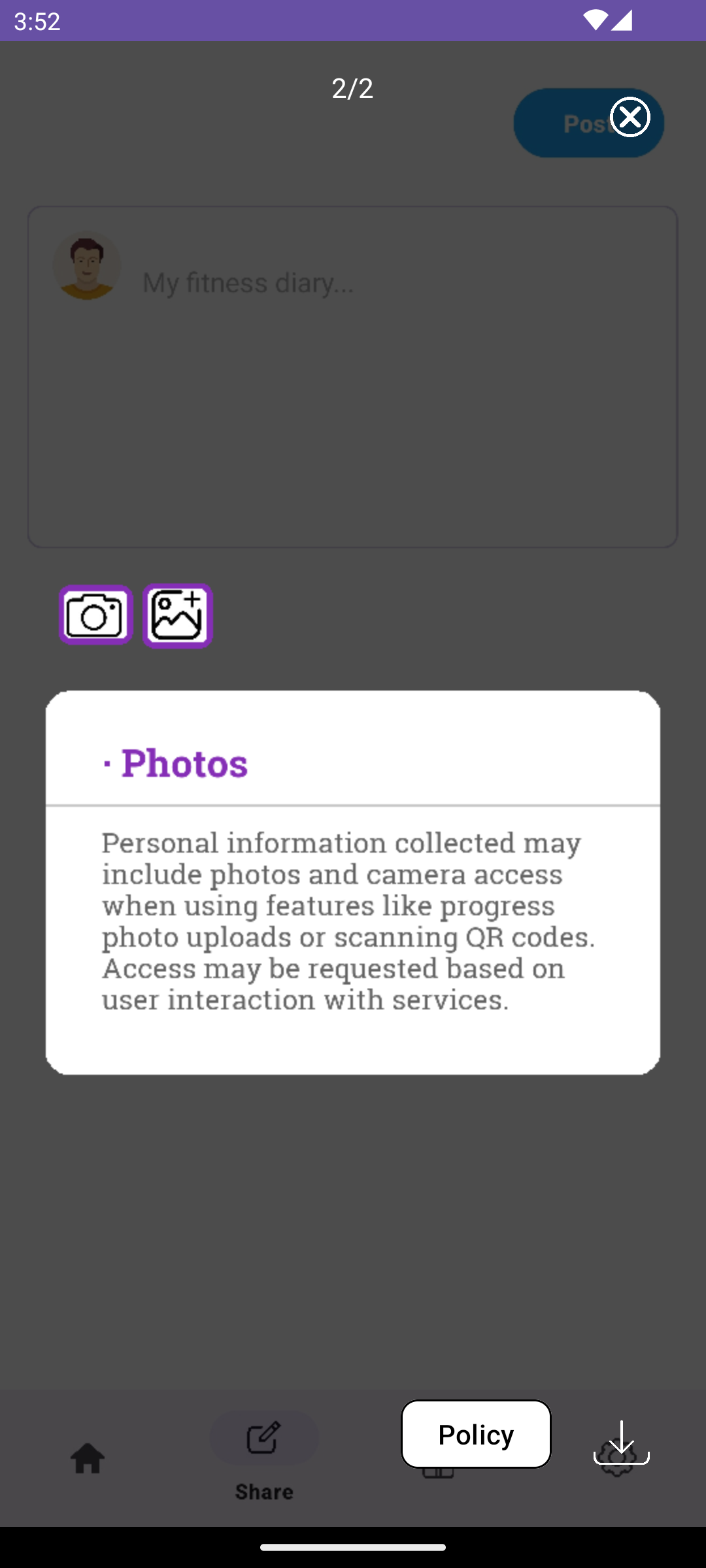}
  \caption{Photos}
  \label{fig_3}
\end{subfigure}
\caption[Caption]{
Figure~\ref{fig_1} displays CPPs for a location icon and text instruction; Figure~\ref{fig_2} presents CPPs for a user profile icon; Figure~\ref{fig_3} shows CPPs for device camera and album icons. 
}
\label{fig_introduction}
\end{figure}

To close this gap, we present the first CPP SDK for mobile applications. Our SDK introduces a lightweight, floating interface button that runs in real time on mobile apps. Upon user interaction, i.e., tapping the button, the system analyzes the current screen, identifies interface elements involved in accessing personal data, and highlights them with overlays. As shown in Fig.~\ref{fig_introduction}, for each identified element, \textbf{\textsc{PrivScan}} presents the associated data type and relevant privacy policy snippets. Developers can easily integrate this SDK into their applications, enabling privacy-aware design and enhancing user trust. 

In Section~\ref{sec_system_description}, we outline the engineering decisions and tool architecture. In Section~\ref{sec_eval}, we present an efficiency evaluation to show the feasibility of \textbf{\textsc{PrivScan}}. The performance and human evaluation of the CPP generation pipeline were conducted in our previous work~\cite{pan2024new}. We conclude in Section~\ref{sec_conclusion}. 

%
%


\section{Tool Architecture}
\label{sec_system_description}

\begin{figure*}[t]
    \renewcommand{\thefigure}{\arabic{figure}}
    \centering
    \includegraphics[width=0.8\linewidth]{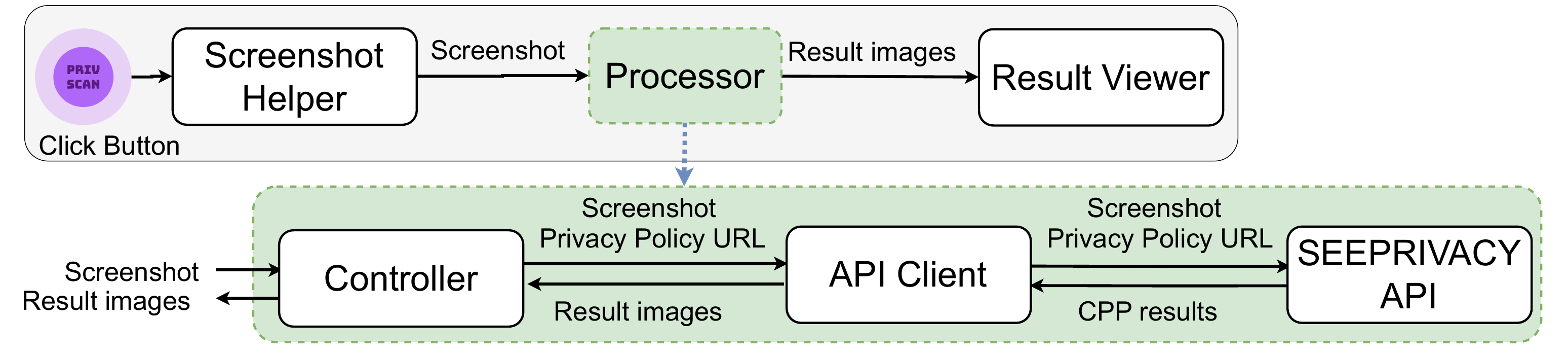}
    \caption{An Overview of \textbf{\textsc{PrivScan}} SDK.}
    \label{fig:sdk}
\end{figure*}


\subsection{Engineering Decisions}

We implement a client-side SDK featuring a floating interaction button, which communicates with the CPP generation framework from our previous work~\cite{pan2024new} via a remotely hosted API. In this section, we introduce the engineering decisions behind the implementation, with consideration of efficiency, user experience and engineering complexity.

\textit{\textbf{Deployment via a Remote API.}}
The \textbf{\textsc{SEEPRIVACY}} pipeline involves both computer vision (CV) and natural language processing (NLP) models that are resource-intensive. On-device execution would therefore incur long inference times and reduced accuracy, especially on mid- to low-end smartphones \cite{ignatov2018ai}. Offloading the pipeline to a server ensures greater stability, delivering uniform performance across heterogeneous hardware. This remote‑API pattern is now a common practice for mobile AI services and simultaneously offers clear cross-platform benefits, where the Python implementation can be reused across different platforms without costly re‑engineering, thereby preserving compatibility with our original research pipeline~\cite{pan2024new} while avoiding a risky full-stack port.


\textit{\textbf{SDK-Based Distribution.}} Distributing \textbf{\textsc{PrivScan}} as a lightweight SDK promotes easy adoption and customization, thereby improving overall developer usability. Developers can integrate the SDK into their apps and tailor the tool to the characteristics of their apps. The SDK also guarantees a seamless in-app user experience, as CPPs appear within the host app. In contrast, a standalone monitoring software would demand continuous background execution and elevated permissions, e.g. overlay windows and usage access, whereas the SDK incurs lower overhead and aligns with the principle of least privilege.



\textit{\textbf{Floating Button Interface.}}
To minimize user effort, we designed a floating interactive button that hovers over the app interface. This widget reduces interaction cost and increases engagement, where a single tap reveals the policy snippet in the corresponding user interface context in a non-disruptive manner. The result is higher usability for app users, consistent with previous findings in mobile user experience that low-effort UI elements increase engagement and task completion, while complex interactions reduce adoption~\cite{priyadarshini2024impact}.


\subsection{Architecture and Workflow}

Below, we present the architecture of \textbf{\textsc{PrivScan}}. As illustrated in Fig.~\ref{fig:sdk}, the tool is composed of five components. On the client side, two frontend modules, Screenshot Helper and Result Viewer are responsible for capturing the GUI screenshots and displaying CPPs, respectively. On the backend side, an API Client handles interactions the remote CPP generation service, and a central Controller orchestrates synchronization among all internal components. Together with the remote API Endpoint, these five components constitute the \textbf{\textsc{PrivScan}} workflow.

\textit{\textbf{Screenshot Helper.}} The Screenshot Helper captures the current user interface of the application. This screenshot serves as one of the two inputs of \textbf{\textsc{PrivScan}}, alongside the privacy policy URL. To eliminate irrelevant visual noise, this module excludes all non-content elements from the screenshot, such as system GUI components, e.g., status bars, and interface elements of \textbf{\textsc{PrivScan}}, e.g., floating buttons. A Bitmap containing an interference-free screenshot of the user interface is created and passed to the Controller and eventually transmitted to the remote server as input to generate CPPs.





\textit{\textbf{Result Viewer.}} The Result Viewer module displays generated CPPs in a manner that is visually coherent and minimally disruptive to the original app interface. The results are rendered as annotated images, each derived from a screenshot of the current interface and leveraging a bounding box to enclose the GUI element that has been identified as related to a specific data type. To avoid overwhelming the user's perception by displaying multiple CPPs on the screen, we create one image per data type and display only one image to the user at a time. When there are elements involving different data on an interface, users can view the CPP of each data type by swiping left or right. We designed a structure‑aware heuristic that adaptively positions CPP boxes, balancing visual association and layout constraints. Instead of fixed placement, the algorithm first extracts all vertical boundaries of annotated icons and interface edges, and computes the gaps between adjacent boundaries. Treating each gap as a candidate region, it selects the tallest gap to minimize overlap and visual clutter. This process yields a consistent, context‑aware display without static templates.
To enhance the readability of CPPs, we leverage a large language model API, e.g., GPT-4o mini~\cite{GPT4omini}, to summarize the original text extracted from the privacy policy to avoid displaying lengthy text directly on the screen.
The currently displayed image can be saved to the device’s media storage when the user presses the save button. The user can exit the viewer at any time by tapping the close icon.

\textit{\textbf{Controller.}} The controller module coordinates other modules in the workflow. Upon initialization, the controller renders a draggable floating button on the screen, allowing users to trigger the CPP generation process within the application.
The button’s position is dynamically adjustable by users, and its size and color can be customized. The controller reacts to a valid click event by inserting a visual mask into the root layout as a full-screen \textit{FrameLayout} with a semitransparent background, temporarily obscuring the underlying interface. At the center of the overlay, a circular \textit{ProgressBar} is presented to inform the user that \textbf{\textsc{PrivScan}} is currently working.  Subsequently, the controller initiates interaction with the Screenshot Helper module to capture the current interface. Once the Screenshot Helper returns a \textit{Bitmap} object representing the captured screen, the controller compresses it into a byte array for network transmission. This byte array, along with the policy URL, serve as inputs required by the CPP generation API. They are submitted through an asynchronous HTTP request dispatched by the API Client to the remote service.

If GUI elements related to data collection are detected on the interface, the controller invokes the result viewer module to render the generated CPPs. Otherwise, a \textit{Toast} message is displayed to notify the user that no element is identified as related to potential privacy data collection. In addition to browsing CPPs, users also have easy access to the application’s full privacy policy through a jump button placed in the lower right corner. In summary, the controller serves as the entry point and central coordinator of the interaction workflow. 

\textit{\textbf{API Client.}} This module mediates data transactions between the controller and the cloud-hosted \textbf{\textsc{SEEPRIVACY}} framework that generates the CPPs. It serves as the primary conduit for uploading inputs and receiving results. This module takes two key inputs: (1) the UI screenshot generated by the Screenshot Helper, and (2) the associated privacy policy URL. 
To initiate the request, the module submits an asynchronous task, constructing a \textit{multipart/form-data} payload and dispatching it to the \textbf{\textsc{SEEPRIVACY}} FastAPI server. 
Once the server completes processing, it returns images with annotated CPPs. Each image corresponds to a data type category identified in the screenshot. 
To accommodate performance and memory constraints on mobile devices, the module employs a two-stage decoding strategy. First, it measures the image dimensions and computes an appropriate scaling factor, e.g., limiting the maximum image dimension to 2,000 pixels. The module then performs actual decoding to construct a \textit{Bitmap} object of the image. After decoding, the module delivers the result images to the controller. 


\textit{\textbf{SEEPRIVACY API.}} The API serves as the backend pipeline for generating CPPs. To support flexible deployment and seamless integration with the mobile SDK, the API is packaged as a standard web service and deployed on the Google Cloud Run platform~\cite{CloudRun}. Our goal is to decouple the image and text processing pipeline from local execution and expose it as a scalable HTTP interface, ensuring consistency while significantly reducing the migration and deployment costs. We present the API implementation in this paper, and more details about the backend pipeline are provided in our previous work~\cite{pan2024new}. The pipeline is containerized using Docker~\cite{docker}. Built upon the \textit{FastAPI} framework~\cite{FastAPI}, the pipeline takes a GUI screenshot and the privacy policy URL as input, which are either retrieved from a caching layer or fetched anew and saved as an HTML file for reuse, reducing response time. This caching strategy also mitigates the risk of anti-bot defenses such as IP blocking triggered by excessive access. 

In summary, through the coordination of the five components (Screenshot Helper, Result Viewer, Controller, API Client, and SEEPRIVACY API), \textbf{\textsc{PrivScan}} integrates contextual privacy policies into practical mobile application scenarios with high modularity.

\section{Evaluation}
\label{sec_eval}

To assess the execution efficiency of the CPP generation pipeline, we conducted an evaluation focusing on the end‑to‑end elapsed time of three key components: Context Detection, Segment Extraction, and CPP Presentation component. For more detailed descriptions of the pipeline components, as well as the performance and human evaluations, please refer to our previous work~\cite{pan2024new}.					
										
We implemented a dummy Android application that integrates our SDK to emulate realistic mobile usage scenarios and invoke the CPP generation pipeline. The dummy app can be run on a Samsung Galaxy S21 5G with 8GB RAM, running Android 13. We used the GPT-4omini~\cite{GPT4omini} to generate a dummy privacy policy for this app.
The app simulates user interactions through privacy-related GUI elements. To cover a variety of interfaces, we designed four application pages: (1) a home page containing both text and icons related to location data collection; (2) a posting page containing icons for profile, device camera, and album; (3) a settings page containing text related to user account; and (4) a rewards page that does not contain any element related to personal data collection. We used first three pages to measure the execution time for each component in the pipeline of CPP generation: Context Detection, Segment Extraction, and CPP Presentation component.

Table~\ref{tab_eval} summarizes the execution times of each component under the three input conditions. Each value is the mean of three runs per screenshot. The overall end-to-end execution time ranges from 8.68\,s to 9.74\,s, indicating that the tool remains within the acceptable latency range of 8-12\,s for complex task in the real world~\cite{shneiderman2010designing}. 
Notably, the Context Detection and Segment Extraction components remain relatively stable across different screenshots, contributing 3.22\,s and 0.69\,s on average, respectively. Context Detection processes the screenshot to identify sensitive UI elements utilizing computer vision techniques, and Segment Extraction extracts the related segments from the privacy policy. A more detailed time measurement of each phase in these components are available at~\url{https://github.com/buyanghc/PrivScan}. CPP Presentation generates the results and shortens the policy segments using LLMs~\cite{GPT4omini}. It accounts for the largest share of the total execution time, with a peak of 5.77\,s under the text-only condition. We attribute this partly to cumulative network variance of the LLM API used to shorten policy text, which reflects real usage, where multiple calls can introduce additional delay. The overall execution time is slightly larger than the sum of the three components as small glue operations between modules were not measured.

Overall, the findings highlight that interface characteristics affect performance. Future optimizations may focus on caching and multimodal prompt engineering to reduce the overhead of CPP generation. Nonetheless, the evaluation confirms that our SDK can run feasibly in mobile apps within the acceptable latency range.


\begin{table}[t!]
\centering
\caption{Execution time per component under three input screenshots: (i) privacy-data–related icons only, (ii) privacy-data–related text only, (iii) mixed icons and text, and the average time cost in seconds.}
\label{tab_eval}
\resizebox{0.99\linewidth}{!}{%
\begin{tabular}{lcccc}
        \toprule
        \textbf{Component} & \textbf{Icon Only (s)} & \textbf{Text Only (s)} & \textbf{Mixed (s)} & \textbf{Average (s)} \\
        \midrule
        Context Detection & 3.55 & 2.92 & 3.20 & 3.22 \\
        Segment Extraction & 0.63 & 0.79 & 0.64 & 0.69\\
        CPP Presentation  & 4.48 & 5.77 & 4.59 & 4.95\\
        \midrule
        \textbf{Overall} & 9.04 & 9.74 & 8.68 & 9.15\\
        \bottomrule
\end{tabular}
}%
\end{table}

\section{Discussion}
\label{sec_discussion}

Numerous usable privacy engineering designs have been proposed to facilitate privacy policy engagement in software engineering and cybersecurity domain, such as policy2label~\cite{pan2023toward}, repo2label~\cite{si2024solution}, CLEAR~\cite{chen2025clear}, and improved App Privacy Report~\cite{wang2025big}.
Despite their promising designs, a persistent limitation hindering their broader adoption is the limited uptake and deployment by developers. 
This can be attributed to challenges such as integration complexity, insufficient developer-centric support, and inadequate usability considerations.
The \textbf{\textsc{PrivScan}} presented in this work exemplifies a practical and usable privacy engineering solution for mobile applications. 
By offering a SDK, the contextual privacy policies can be seamless integrated into mobile applications, thereby lowering the barrier and additional engineering effort for adoption by developers.

Thus, we further advocate for increased research attention towards the implementation aspect that not only promote the adoption of usable privacy designs by developers but also minimize their burden. 
Fostering developer-centric solutions is critical to bridging the gap between privacy research and real-world deployment.

\section{Conclusion and Future Work}
\label{sec_conclusion}

Concerns regarding data collection practices of mobile applications are growing, while privacy policies remain lengthy and disconnected from real-time usage contexts, making it difficult for users to understand how their data is collected and used. In this work, we extend the framework in our previous work~\cite{pan2024new} with a deployable SDK, \textbf{\textsc{PrivScan}}, which can be easily integrated into mobile apps and provides real-time, non-intrusive contextual privacy policies via a floating interactive button. We believe \textbf{\textsc{PrivScan}} illustrates the practicality of CPPs and is a step towards its broader real‑world adoption. 

Future work will focus on refining the system's user experience through iterative feedback and usability testing. 
A key direction involves exploring dynamic privacy policy segmentation strategies that tailor content not only by data type, but also according to individual user privacy preferences, thereby enhancing the contextual relevance and personalization of privacy communications.
Additional avenues include evaluating long-term user engagement, incorporating adaptive policy presentation techniques, and extending the system for cross-platform deployment to further improve accessibility and promote wider adoption.



\bibliographystyle{IEEEtran}
\bibliography{References}

\end{document}